\documentclass[12pt,preprint]{aastex}

\shorttitle{Radio Emission from IMBH in G1}
\shortauthors{Ulvestad et al.}
\def\H2{\ion{H}{2}}

\begin{document}

\title{Radio Emission from the Intermediate-mass Black Hole in 
the Globular Cluster G1}

\author{James S.~Ulvestad}
\affil{National Radio Astronomy Observatory}
\affil{P.O. Box O, Socorro, NM 87801}
\email{julvesta@nrao.edu}
\author{Jenny E.~Greene$^1$}
\affil{Department of Astronomy, Princeton University}
\affil{Princeton, NJ}
\email{jgreene@astro.princeton.edu}
\altaffiltext{1}{Hubble Fellow}
\and
\author{Luis C.~Ho}
\affil{The Observatories of the Carnegie Institute of Washington}
\affil{813 Santa Barbara St., Pasadena, CA 91101}
\email{lho@ociw.edu}
\setcounter{footnote}{1}

\begin{abstract}

We have used the Very Large Array (VLA) to search for radio emission from
the globular cluster G1 (Mayall-II) in M31. G1
has been reported by Gebhardt et al. to contain an intermediate-mass
black hole (IMBH) with a mass of $\sim 2\times 10^4~M_\odot$.
Radio emission was detected within an arcsecond of the cluster
center with an 8.4 GHz power of $2\times 10^{15}$~W~Hz$^{-1}$.
The radio/X-ray ratio of G1 is a few hundred times
higher than that expected for a high-mass X-ray binary in the
cluster center, but is consistent with the expected value 
for accretion onto an IMBH with the reported mass.
A pulsar wind nebula is also a possible candidate for the radio and
X-ray emission from G1; future high-sensitivity VLBI observations
might distinguish between this possibility and an IMBH.
If the radio source is an IMBH, and similar accretion and
outflow processes occur for hypothesized 
$\sim 1000~M_\odot$ black holes in Milky Way globular clusters,
they are within reach of the current VLA and should be detectable easily
by the Expanded VLA when it comes on line in 2010.

\end{abstract}


\section{Introduction}
\label{sec:intro}

Most galaxies with massive spheroidal components appear to harbor central 
black holes (BHs), with masses ranging from a few $\times 10^6~M_\odot$ to 
over $10^9~M_\odot$.  These BH masses are
well correlated with both the luminosity and the velocity dispersion
of the galaxy spheroid \citep{geb00,fer00}, implying that the
formation of the central BHs is connected intimately with the
development of the galaxy bulges.  However, a bulge is not necessarily
a prerequisite for a massive BH.  On the one hand, neither the
late-type spiral galaxy M33 (Gebhardt et al. 2001) nor the dwarf
spheroidal galaxy NGC 205 (Valluri et al. 2005) shows dynamical
evidence for a massive BH.  On the other hand, both the Sd
spiral galaxy NGC 4395 and the dwarf spheroidal galaxy POX 52 contain
active BHs with masses $\approx 10^5~M_\odot$, although
neither contains a classic bulge (Fillipenko \& Ho 2003; Barth et
al. 2004).  Greene \& Ho (2004, 2007) find that optically active
intermediate-mass BHs (IMBHs, $M_{\rm BH} < 10^6~M_\odot$), while rare,
do exist in dwarf galaxies, but optical searches are heavily biased toward 
sources accreting at high Eddington rates.  Alternate search techniques are 
needed to probe the full demographics of IMBHs.

Direct dynamical detection of IMBHs is currently impossible outside of the 
Local Group.  However, Gebhardt, Rich, \& Ho (2002, 2005) have found dynamical 
evidence for an excess dark mass of $(1.8\pm 0.5)\times 10^4~M_\odot$ at the 
center of the globular cluster G1 in M31; the evidence for this IMBH was
questioned by \citet{bau03}, but supported by the improved data and
analysis of Gebhardt et al. (2005).  The physical nature of the 
central dark object is difficult to prove: it could be either an IMBH or a 
cluster of stellar remnants.  While the putative presence of 
a BH in a globular cluster center may appear unrelated to galaxy 
bulges, the properties of G1, including its large mass, high degree of 
rotational support, and multi-aged stellar populations, all suggest that G1 is
actually the nucleus of a stripped dwarf galaxy \citep{mey01}.  Most
intriguingly, the inferred BH mass for G1 is about 0.1\% of
the total mass, consistent with the relation seen for higher mass BHs, 
and consistent with predictions based on mergers of BHs
\citep{mil02} or stellar mergers in dense clusters \citep{por02}.

Recently, \citet{poo06} reported an X-ray detection of G1, with a 0.2--10~keV 
luminosity of $L_{\rm X}\approx 2\times 10^{36}$~ergs~s$^{-1}$.  Although
this may represent accretion onto a central BH, it is within the 
range expected for either accretion onto an IMBH or for a massive X-ray
binary.  Unfortunately, the most accurate X-ray position determined recently 
by \citet{kon07} does not have sufficient accuracy to determine whether the
X-ray source is located within the central core of G1, which would
help distinguish between these two possibilities.  The radio/X-ray
ratio of G1 provides an additional test of the nature of the G1 X-ray
source.  As pointed out by \citet{mac04} and Maccarone, Fender, \&
Tzioumis (2005), deep radio
searches may be a very effective way to detect IMBHs in globular
clusters and related objects, since, for a given X-ray luminosity,
stellar mass BHs produce far less radio luminosity than supermassive
BHs.  The relation between BH mass, and X-ray and radio luminosity
empirically appears to follow a ``fundamental plane,'' in which the
ratio of radio to X-ray luminosity increases as the $\sim 0.8$ power
of the BH mass (Merloni, Heinz, \& di Matteo 2003; Falcke, K\"ording,
\& Markoff 2004).  For an IMBH mass of
$1.8\times 10^4~M_\odot$ in G1, one thus would expect a radio/X-ray
ratio about 400 times higher than for a $10~M_\odot$ stellar BH.
In this paper, we report a deep Very Large Array (VLA)
integration on G1 and a radio detection that apparently confirms the
presence of an IMBH whose mass is consistent with that found by
Gebhardt et al. (2002, 2005).

\section{Observations and Imaging}
\label{sec:data}

We obtained a 20-hr observation of G1 using the VLA in its
C configuration (maximum baseline length of 3.5~km) 
at 8.46 GHz.  The observation was split into two 10-hr sessions, one 
each on 2006 November 24/25 and 2006 November 25/26.
Each day's observation consisted of repeated cycles of 1.4
minutes observation on the local phase calibrator J0038+4137
and 6 minutes observation on the target source G1.  In addition,
each day contained two short observations of 3C~48 (J0137+3309)
that were used to calibrate the flux density scale to
that of \citet{baa77}.  Thus, the total integration time on G1 was
14.1 hr.  We also obtained a total of 9.5 hr of observing in C 
configuration at 4.86~GHz on 2007 January 13/14 and 2007 January 14/15, 
using a similar observing strategy, and achieving
a total of 7.3 hr of integration on source.

All data calibration was carried out in NRAO's Astronomical Image 
Processing System \citep{gre03}.  Absolute antenna gains
were determined by the 3C~48 observations, then transferred to
J0038+4137, which was found to have respective flux densities of 
0.52~mJy and 0.53~mJy at 8.4 and 4.9~GHz.  In turn, J0038+4137 
was used to calibrate the interferometer amplitudes
and phases for the target source, G1.  Erroneous data were 
flagged by using consistency of the gain solutions as a guide
and by discarding outlying amplitude points.

The VLA presently is being replaced gradually by the Expanded VLA
(EVLA), which includes complete replacement of virtually all the
electronic systems on the telescopes.  Since antennas are
refurbished one at a time, the VLA at the time of our observations
consisted of 18--20 ``old'' VLA antennas and 6 ``new'' (actually,
refurbished) EVLA antennas, having completely different electronics
systems.  Although all antennas were cross-correlated for our
observations, we found subtle errors in some of the EVLA data.  
Thus, to be conservative, we discarded the data from all EVLA
antennas except for 3 antennas that were confirmed to work
very well on 2006 November 24/25.

The radio data were Fourier transformed 
and total-intensity images were produced in each band, covering
areas of 17\arcmin$\times$17\arcmin\ at each frequency.\footnote{The
image areas covered were much larger than the primary beams of
the individual VLA antennas, in order to provide the best
possible subtraction of confusing sources.} These images were CLEANed 
in order to produce the final images.  At 8.4~GHz, the rms noise was 
6.2~$\mu$Jy~beam$^{-1}$ for a beam size of 2\farcs94$\times$2\farcs72;
at 4.9~GHz, the noise was 15.0~$\mu$Jy~beam$^{-1}$ for a beam
size of 5\farcs09$\times$4\farcs43.

A few radio sources with strengths of hundreds of microjansky to
a few millijansky were found in the images, but we discuss
only G1 in this {\it Letter}.\footnote{The strongest nearby radio source 
is a 1~mJy (at 8.4~GHz) object located approximately 100\arcsec\ from
G1; this object and the other weak sources might be X-ray binaries or
supernova remnants if located at the distance of M31, but their numbers
also are consistent with the possibility that some could be background
extragalactic sources.}  At 8.4~GHz,
an apparent source with a flux density of $28\pm 6$~$\mu$Jy
[corresponding to $2\times 10^{15}$~W~Hz$^{-1}$ for 
distance modulus of $(m-M)=24.42$~mag (Meylan et al. 2001)]
was found approximately one arcsecond from the G1 optical
position reported by \citet{mey01}; this radio source has
J2000 coordinates of $\alpha=00^h32^m46.54^s$, 
$\delta=39^\circ 34'39.2''$.  Figure~1 shows our
8.4~GHz image of the 20\arcsec\ by 20\arcsec\ region centered
on G1; this image includes a $1~sigma$ error circle of
1\farcs5 radius for the X-ray position found by \citet{kon07}.
The radio position has an estimated error of 0\farcs6 in
each dimension (not shown in the figure), derived by dividing the 
beam size by the signal-to-noise ratio.\footnote{Transfer of the 
phase from the local phase calibrator makes an insignificant 
contribution to the radio source position error, relative to the
error imposed by the limited signal strength.}

The {\it a priori} probability of finding a $4.5~\sigma$ noise spike or 
background source so close to G1 is quite small, as indicated by the lack 
of any other contours of similar strength in Figure~1.  If we hypothesize 
that there are 9 independent beams (roughly 8\arcsec\ by 8\arcsec) within 
which a source would be considered to be associated with G1, then
the probability of a $4.5~\sigma$ noise point close to G1 is less
than $10^{-4}$.  Similarly, the expected density of extragalactic
radio sources at 28~$\mu$Jy or above is 0.25 arcmin$^{-2}$
\citep{win93}, or $4\times 10^{-3}$ in a box 8\arcsec\ on a side,
 making it unlikely that we have found an unrelated
background source.  In order to search for possible data errors
that might cause a spurious source, we have subjected 
our data set to additional tests, imaging data from the two days
separately, and also imaging the two different intermediate 
frequency channels separately.  The G1 radio source remains 
in the images made from each data subset, with approximately the 
same flux density and position. The overall significance is reduced 
by $2^{1/2}$ to approximately $3~\sigma$ in each image made with about 
half the data, as expected for a real source with uncontaminated
data.  Other $2.5~\sigma$--$3~\sigma$ sources appear in the central 
20\arcsec\ box in some subsets of half the data, consistent with 
noise statistics, but none is above the $3.5~\sigma$
level in the full data set.  Thus, all tests indicate that the
detection of G1 is real, and we will proceed on that basis
for the remainder of this paper.  At 4.9~GHz, we find no detection
at the G1 position, but the much higher noise level provides us
only with very loose constraints on the source spectrum (see below).

\section{Origin of the G1 Radio Emission}

\citet{mer03} and \citet{fal04} have quantified an empirical
relation (or ``fundamental plane'') among X-ray and 5~GHz radio 
luminosity and BH mass; we use
the \citet{mer03} relation $L_R\propto L_{\rm X}^{0.6}M_{\rm BH}^{0.78}$.  
\citet{mer03} analyzed this relation in the context of accretion
flows and jets associated with massive BHs.  One might expect
some general relationship among these three quantities, if 
an X-ray-emitting accretion flow onto a massive BH leads to
creation of a synchrotron-emitting radio jet, with the detailed
correlation providing some insight into the nature of that flow.
By comparing the empirically determined relation with expectations
from theoretical models, \citet{mer03} deduced that the data for
BHs emitting at only a few percent of the Eddington rate 
are consistent with radiatively inefficient accretion flows and a 
synchrotron jet, but inconsistent with standard disk accretion models.

\citet{mac04} scaled the fundamental-plane relation to values appropriate 
for an IMBH in a Galactic globular cluster; we rescale their equation
here to find a predicted radio flux density of

\begin{equation}
S_{\rm 5\ GHz}\ =\ 52\ \biggl({L_{\rm X}\over {10^{36}\ {\rm ergs\ s}^{-1}}}
\biggr)^{0.6}\ \biggl({M_{\rm BH} \over 10^4~M_\odot}\biggr)^{0.78}\ 
\biggl({d\over 600\ {\rm kpc}}\biggr)^{-2}\ \mu{\rm Jy}\ .
\end{equation}

Using the previously cited X-ray luminosity and BH mass for G1 and our 
adopted distance modulus,
this predicts a 5~GHz
flux density of 77~$\mu$Jy for G1.  However, taking into account
the 30\% uncertainty in the IMBH mass, the unknown spectral
index of the radio emission, and the dispersion of 0.88 in
$\log L_R$ \citep{mer03}, the predicted 8.4~GHz flux density for G1 
is in the range of tens to a few 
hundred microjansky.  Thus, our radio detection of 28~$\mu$Jy at 
8.4~GHz is consistent with the predictions for a 
$1.8\times 10^4~M_\odot$ IMBH, but strongly inconsistent with a 
$10~M_\odot$ BH.  Since neutron star X-ray binaries in a
variety of states have radio/X-ray ratios much lower than 
BH X-ray binaries \citep{mig06}, and thus another 2 orders
of magnitude below the observed value, 
stellar-mass X-ray binaries of any type are ruled out as the possible
origin of the radio emission in G1.

We can use the radio/X-ray
ratio to assess other possible origins for the radio emission.  Here, we
use the ratio $R_{\rm X} = \nu L_\nu({\rm 8.4\ GHz})/L_{\rm X}({\rm 2-10\ keV)}$
as a fiducial marker. For G1, $R_{\rm X}\approx 5\times 10^{-5}$, which is
considerably lower than $R_{\rm X} \approx 10^{-2}$ that is common to
the Galactic supernova remnant Cas~A, low-luminosity active
galactic nuclei (supposing G1 might be a stripped dwarf elliptical
galaxy), and most ultraluminous X-ray sources (cf. Table~2 of Neff, 
Ulvestad, \& Campion [2003], and references therein).  

It is of interest to compare the G1 source to various relatives
of pulsars as well.  For
instance, G1 is within the wide range of both luminosity and radio/X-ray 
ratio observed for pulsar wind nebulae (PWNs) \citep{fra97}, less luminous
than the putative PWN in M81 \citep{bie04}, but
considerably more luminous than standard pulsars or anomalous X-ray
pulsars \citep{hal05}.  The 8.4~GHz luminosity of G1 is similar to that
of the magnetar SGR~$1806-20$ about 10 days after its outburst in late 2004,
and the lack of a 4.9~GHz detection would be consistent with
the fading of SGR~$1806-20$ two months after the outburst
\citep{gae05}.  However, there is no published evidence for a gamma-ray
outburst from G1, and the relatively steady apparent X-ray flux
\citep{poo06} also argues against a transient source.  Thus, the only
stellar-mass object that might account for the radio and X-ray emission
would be a PWN; using the scaling law given by
\citet{fra97}, we find a likely size of $\sim 10$ milliarcseconds
for a PWN radio source in G1, implying that high-sensitivity VLBI
observations could distinguish between a PWN and IMBH origin
for the radio emission from G1.

Knowledge of the radio spectrum of G1 could provide more clues to the
character of the radio emission, although either a PWN 
or an IMBH accretion flow might have a flat spectrum.
In any case, our 5~GHz observation simply is not deep enough.  
If we choose a $2~\sigma$ upper limit of 30.0~$\mu$Jy at 4.9~GHz 
($2~\sigma$ chosen since we know the position of the 8.4~GHz source with 
high accuracy), we derive a spectral index limit of $\alpha > -0.12 \pm 0.99$ 
(for $S_\nu\propto \nu^{+\alpha}$, $1~\sigma$ error in spectral index), 
which has little power to discriminate among models.

The X-ray emission from G1 may be due to 
Bondi accretion on the IMBH, either from ambient cluster gas or from
stellar winds \citep{poo06}.  \citet{ho03} and \citet{poo06} give 
approximate relations for the Bondi accretion on an IMBH in a globular
cluster; for an ambient density of 0.1~cm$^{-3}$, an ambient speed
of 15~km~s$^{-1}$ for the gas particles relative to the IMBH, and a 
radiative efficiency of 10\%, the Bondi accretion luminosity for
the G1 IMBH would be $\sim 3\times 10^{38}$~ergs~s$^{-1}$.  The X-ray
luminosity of $2\times 10^{36}$~ergs~s$^{-1}$ measured by \citet{poo06} 
thus implies accretion at just under 1\% of the Bondi rate.  Given that 
$L_{\rm X}/L_{\rm Edd} \approx 10^{-6}$, a more likely scenario is 
that G1 accretes at closer to 10\% of the Bondi rate but with a 
radiative efficiency under 1\%.  In this context, we
note that the radio/X-ray ratio for G1 is $\log R_{\rm X} > -4.3$, which is 
above the value of $-4.5$ used to divide radio-quiet from radio-loud
objects \citep{ter03}.\footnote{A lower limit is given for $R_{\rm X}$
because this quantity traditionally is given in terms of the 2--10~keV
luminosity, whereas $\log R_{\rm X}=-4.3$ would correspond to the 
value computed for the 0.2--10~keV luminosity given by \citet{poo06}.} 
G1 therefore should be considered radio-loud,
as inferred for BHs in galactic nuclei that radiate well below
their Eddington luminosities (Ho 2002).

If the globular clusters in our own Galaxy also have central BHs
that are 0.1\% of their total masses, and they accrete and radiate in the
same way as G1, many would have expected 5~GHz radio flux densities 
in the 20--100~$\mu$Jy range; flux densities often would be in the
1--10~$\mu$Jy range even for less efficient accretion and 
radiation \citep{mac05}.  As \citet{mac05} summarize,
there are few radio images of globular clusters that go deep enough
to test this possibility.  \citet{fen04} points out that the Square
Kilometer Array (SKA) will be able to test for the existence of IMBHs in
many globular clusters.  However, based on our results for G1, we suggest
that it is not necessary to wait for the SKA; the current VLA can reach
the hypothesized flux densities with some effort.  The EVLA 
\citep{ulv06}, scheduled to be on line in about 2010, will have 40 times
the bandwidth and 6.3 times the sensitivity of the current VLA in
the frequency range near 8~GHz.  This will enable the EVLA to reach
the 1~$\mu$Jy noise level in approximately 12 hours of integration,
thus probing the range of radio emission predicted by \citet{mac05}
for many globular clusters.

\section{Summary}

We have detected faint radio emission from the object G1, a globular 
cluster or stripped dwarf elliptical galaxy in M31.  The emission has
an 8.4~GHz power of $2\times 10^{15}$~W~Hz$^{-1}$.  Assuming that the
radio source is associated with the X-ray source in G1 \citep{poo06},
the radio/X-ray ratio is consistent with the value expected for an
accreting $\sim 2\times 10^4~M_\odot$ BH. 
Thus, the radio detection lends support to the
presence of such an IMBH within G1.  The
other possible explanation, a pulsar wind nebula, could be tested
by making very high-sensitivity VLBI observations of G1.

\acknowledgments

The National Radio Astronomy Observatory is a facility of the 
National Science Foundation operated under cooperative agreement by Associated 
Universities, Inc.  We thank the staff of the VLA that made these 
observations possible.  Support for JEG was provided by NASA through
Hubble Fellowship grant HF-01196, and LCH acknowledges support from NASA grant 
HST-GO-09767.02.  Both were awarded by the Space Telescope Science 
Institute, which is operated by the Association of Universities for
Research in Astronomy, Inc., for NASA, under contract NAS~5-26555.
We also thank Dale Frail for useful discussions about pulsar wind
nebulae, and an anonymous referee for useful suggestions.

{\it Facilities:} \facility{VLA}.

\clearpage

\begin{figure}
\plotone{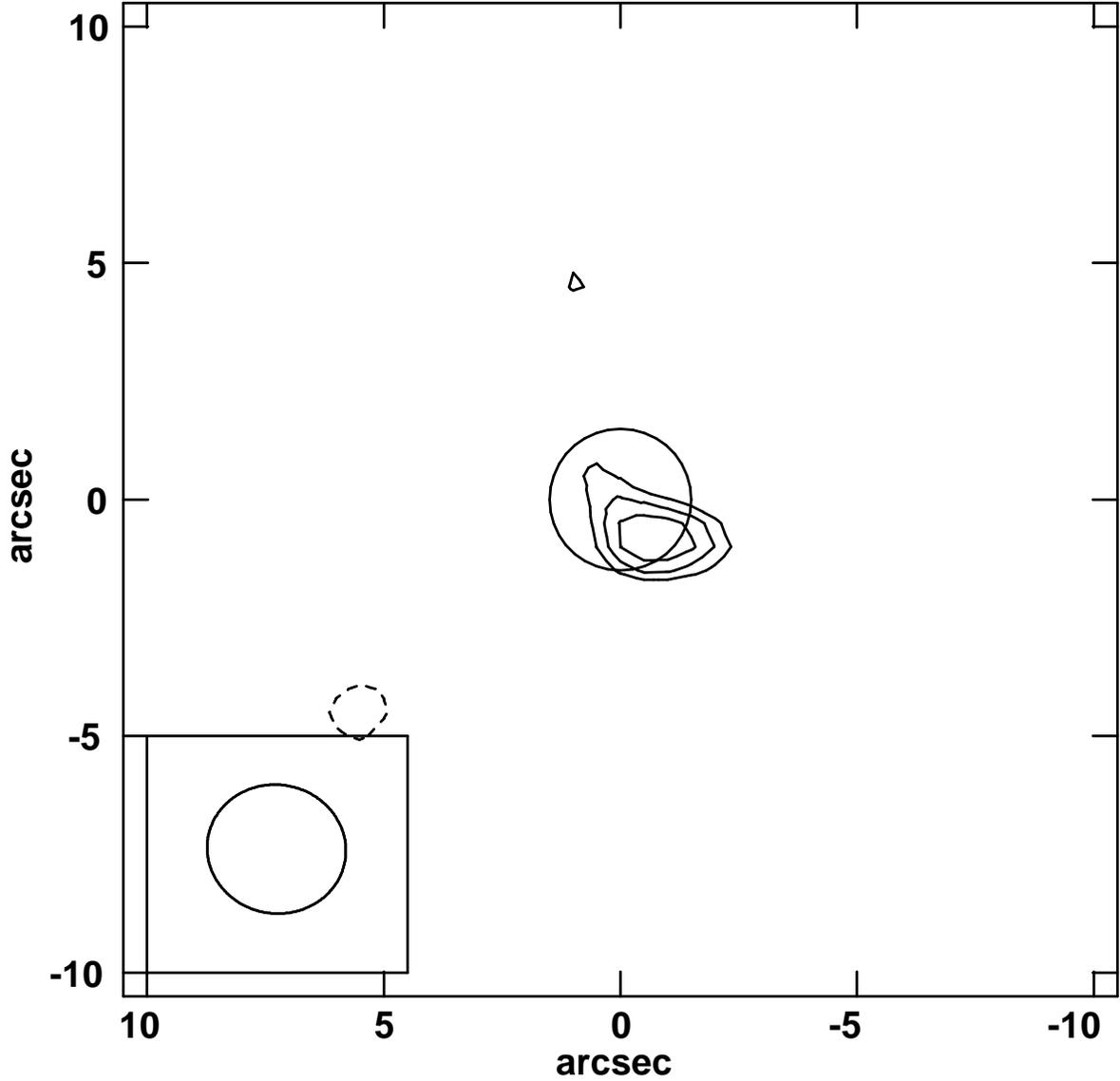}
\caption{
VLA C configuration 8.4~GHz image of the vicinity of G1 in M31, with
contours at intervals of 0.5 times the r.m.s noise.  The lowest
contour is at 3 times the noise of 6.2~$\mu$Jy~beam$^{-1}$, and
negative contours are shown dashed; the synthesized beam is shown
in the box in the lower-left corner.  The (0,0) point is at a
J2000 position of $\alpha=00^h32^m46.60^s$, $\delta=39^\circ 34'40.0''$.
The 1\farcs5-radius circle about this central point represents
the $1~\sigma$ error in the X-ray position of G1 \citep{kon07}. }
\label{fig:g1xband}
\end{figure}

\clearpage


\end{document}